\pdfoutput=1
\documentclass[pdflatex, sn-mathphys]{sn-jnl}
\jyear{2023}
\usepackage{etoolbox}
\usepackage{indentfirst}
\theoremstyle{thmstyleone}

\theoremstyle{thmstyletwo}

\theoremstyle{thmstylethree}

\makeatletter
\patchcmd{\ps@headings}
{\hbox to \hsize{\hfill Springer Nature 2021 \LaTeX\ template\hfill}}
{\hbox to \hsize{}}
{}
{}
\patchcmd{\ps@headings}
{\hbox to \hsize{\hfill Springer Nature 2021 \LaTeX\ template\hfill}}
{\hbox to \hsize{}}
{}
{}
\patchcmd{\ps@titlepage}
{\hbox to \hsize{\hfill Springer Nature 2021 \LaTeX\ template\hfill}}
{\hbox to \hsize{}}
{}
{}
\makeatother

\begin{document}

\title[ ]{Imaging the diffusive-to-ballistic crossover of magnetotransport in graphene}

\author[1]{\fnm{Zachary J.} \sur{Krebs}}

\author[2]{\fnm{Wyatt A.} \sur{Behn}}

\author[1]{\fnm{Keenan J.} \sur{Smith}}

\author[1]{\fnm{Margaret A.} \sur{Fortman}}

\author[3]{\fnm{Kenji} \sur{Watanabe}}

\author[4]{\fnm{Takashi} \sur{Taniguchi}}

\author[5]{\fnm{Pathak S.} \sur{Parashar}}

\author[5]{\fnm{Michael M.} \sur{Fogler}}

\author[1]{\fnm{Victor W.} \sur{Brar}}\email{vbrar@wisc.edu}

\affil[1]{\orgdiv{Department of Physics}, \orgname{University of Wisconsin - Madison}, \orgaddress{\city{Madison}, \state{WI} \postcode{53706}, \country{USA}}}

\affil[2]{\orgdiv{Department of Physics}, \orgname{McGill University}, \orgaddress{\city{Montréal}, \state{QC}, \country{Canada}}}

\affil[3]{\orgdiv{Research Center for Electronic and Optical Materials}, \orgname{National Institute for Materials Science}, \orgaddress{\city{Tsukuba} \postcode{305-0044}, \country{Japan}}}

\affil[4]{\orgdiv{Research Center for Materials Nanoarchitectonics}, \orgname{National Institute for Materials Science}, \orgaddress{\city{Tsukuba} \postcode{305-0044}, \country{Japan}}}

\affil[5]{\orgdiv{Department of Physics}, \orgname{University of California - San Diego}, \orgaddress{\city{La Jolla}, \state{CA} \postcode{92093}, \country{USA}}}


\abstract{Scanning tunneling potentiometry (STP) is used to probe the local, current-induced electrochemical potential of carriers in graphene near circular electrostatic barriers in an out-of-plane magnetic field ranging from 0 to 1.4 T. These measurements provide nanometer-resolved information about the local motion of carriers, revealing significant changes in carrier dynamics with increasing field strength. At low magnetic fields the electrochemical potential displays a spiral-like pattern, while at high fields it exhibits distinct changes at particular radii. We show that the observed behavior indicates a transition from diffusive to ballistic transport. Additionally, the sharp changes in the measured potential profile at high fields result from the `spirograph' motion of carriers, which creates a local enhancement of the Hall field one cyclotron diameter away from the semiclassical turning point near the electrostatic barrier.} 

\maketitle

At low temperatures and strong magnetic fields, the quantum Hall effect (QHE) can be observed in two-dimensional materials by measuring the current-induced voltage drop between terminals placed along the sample edges \cite{klitzing1980}. Similar `Hall bar' experiments over the years have uncovered a wealth of DC magnetotransport phenomena besides the QHE, including non-equilibrium resistance oscillations as well as ballistic and hydrodynamic effects \cite{yang2002, gusev2018, greenaway2021, wang2023}. These changes to the macroscopic resistivity tensor arise from the suppression (or enhancement) of different electron scattering channels, with changing temperature, field strength, and carrier density. The specific hierarchy of electron scattering rates in a material can also affect the microscopic flow of current around obstacles and disorder \cite{guo2017, jenkins2022, krebs2023}. In the ballistic transport regime, nonlocal correlations related to the cyclotron diameter are expected to disturb the usual diffusive flow of current, generating nanometer-scale peaks in the Hall field that are inaccessible to standard resistance measurements \cite{holder2019, raichev2020}. Detecting these local transport signatures requires noninvasive probes with high spatial and energy resolution,  beyond that of electro-optic techniques \cite{fontein1991, knott1995, cui2016}. Directly visualizing current flow and its associated Hall potential at the nanometer scale, across a wide range of magnetic fields and in the presence of known disorder potentials, is within reach of scanned probes that function as local ammeters or potentiometers.

Previously, atomic force microscopy (AFM) has been effective in spatially mapping the Hall potential of 2D electron systems confined to GaAs heterojunctions, especially in the quantum Hall regime \cite{mccormick1999, weitz2000, weis2011, ahlswede2001}. For dissipationless transport, the Hall potential drop was reported to coincide with the positions of current-carrying incompressible strips. Ultimately, the limited resolution of AFM in these experiments prevented local potential imaging near individual defects, charge disorder, and compressible droplets. In particular, AFM cantilevers are known to feel a weighted average of all electrostatic forces acting on the tip -- not just those induced by a source-drain bias -- which lowers both the spatial and energy resolution in a  way that depends on the detailed tip geometry \cite{glatzel2003, behn2021}. Despite this challenge, a recent AFM experiment imaged broken-symmetry quantum Hall edge channels in graphene along a strong confining gate potential \cite{kim2021}.  

Besides potentiometry with an AFM tip, such as scanned single-electron transistors \cite{yacoby1999, ilani2004}, local capacitance imaging \cite{tessmer1998, finkelstein2000, suddards2012} and scanning tunneling microscopy \cite{hashimoto2008, li2013} have also been used to probe quantum Hall states. These techniques were successful in resolving sub-micron features associated with the QHE, including topological edge modes and localized bulk states. Recently, the first direct detection of equilibrium current carried by individual edge modes in graphene was achieved by means of scanned superconducting quantum interference device \cite{uri2020}. These higher-resolution approaches typically focus on the equilibrium properties of quantum Hall states, and do not directly access information about the Hall field, electron scattering processes, and current distribution induced by a source-drain bias. Methods that rely on a transport current, such as scanning gate microscopy and nanothermometry, have tested the robustness of the QHE against tunable disorder as well as the topological protection of edge modes against various scattering channels \cite{baumgartner2007, marguerite2019, wang2021}. 

Scanning tunneling potentiometry (STP) is an alternative approach that uses a standard STM tip to spatially map the electrochemical potential ($\mu_\mathrm{ec}$) of a surface induced by an applied source-drain bias in a thin conductor \cite{muralt1986}.  STP can provide sub-\AA \ resolved images of $\mu_\mathrm{ec}$ under transport conditions, which in turn may reveal key aspects of the electronic flow \cite{briner1996, krebs2023}. With the ultra-high mobility graphene samples of today, STP is well-suited to visualize nanoscale magnetotransport from zero field all the way through the quantum Hall regime. To this end, Willke et. al. used STP to spatially map the bulk Hall potential drop across graphene/SiC samples in magnetic fields up to 6 T, correlating sharp drops in the STP signal with individual step edges that scatter incident charge carriers \cite{willke2017}. Those results could be explained in terms of purely diffusive (Ohmic) magnetotransport without ballistic or quantum corrections.

In this work, we use STP to image carrier flow around electrostatic barriers in graphene/hBN samples with large momentum-relaxing mean free paths ($l_\mathrm{mr} > 2$ $\mu$m)  for charge carriers at low temperature ($T = 4.5$ K). As we increase the magnetic field past 0.6 T, we observe a diffusive-to-ballistic crossover in which the guiding center drift of carriers along equipotential lines is no longer suppressed by scattering. The measured deviations from purely Ohmic transport at high magnetic fields can be modeled numerically with tight-binding simulations, and explained analytically with solutions to the Boltzmann equation in an appropriate limit. Both the numerical and analytical methods predict a local enhancement of the Hall field one cyclotron diameter ($2 R_\mathrm{c}$) away from the barrier  \cite{holder2019, raichev2020}. This local enhancement appears in the STP data, constituting a direct measurement of cyclotron motion and finite size effects in 2D electron systems.

\begin{figure}
\centering
\includegraphics[width=\textwidth,keepaspectratio]{"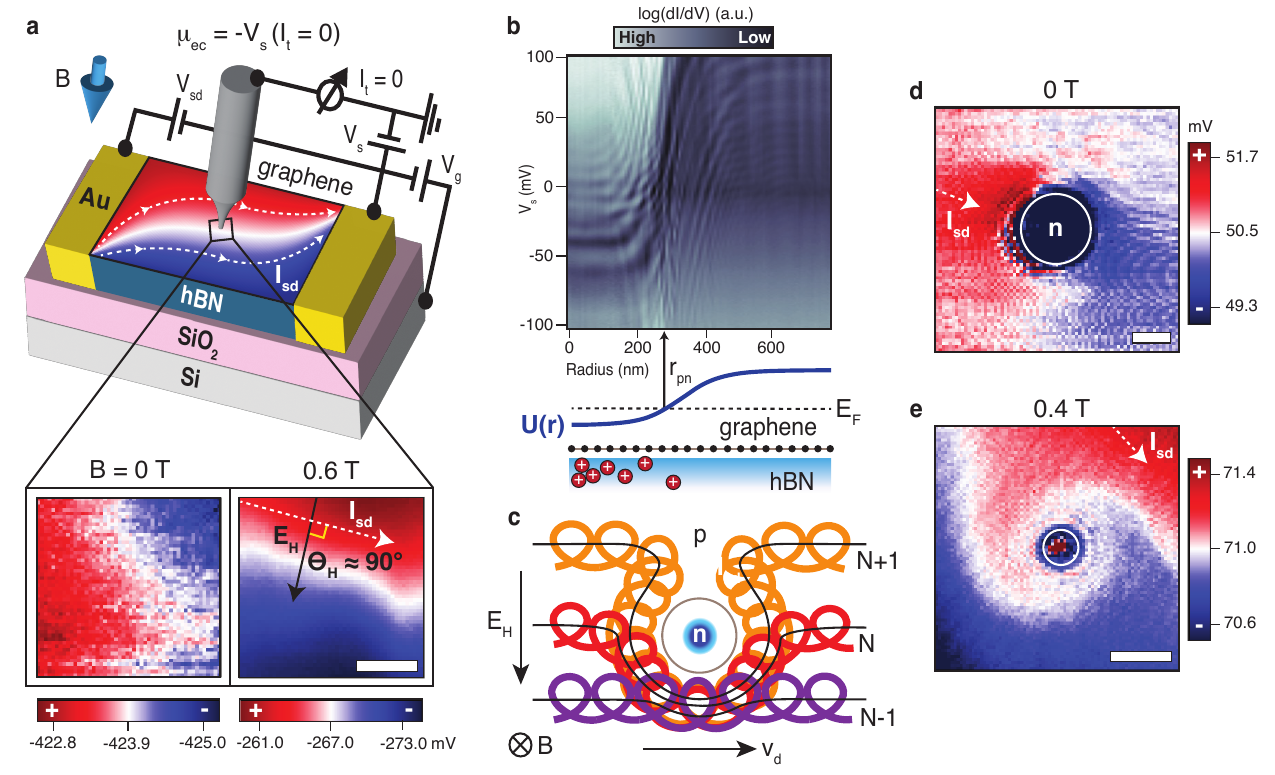"}
\caption{(a) Experimental setup of the STP measurement. An STM tip raster-scans across the graphene sheet and records the electrochemical potential $\mu_\mathrm{ec}$ induced by a source-drain bias. The dashed white arrows indicate the direction of current flow. In a micron-scale patch of the graphene sheet, $\mu_\mathrm{ec}$ resembles a tilted plane (shown for two magnetic fields, $B = 0$ and $0.6$ T, at $I_\mathrm{sd} = 100$ $\mu$A). The scale bar is 500 nm. (b)-(c) Effect of a circular p-n junction on the electronic structure and charge carrier motion in the same scan window. (b) STS measurements along a radial linecut through the center of the p-n junction with $B = 1.4$ T and $V_\mathrm{g} = -30$ V. A cartoon shows the electrostatic potential $U(r)$ (blue line) generated by ionized defects in the hBN substrate. The intersection of the Fermi level, $E_\mathrm{F}$, with $U(r)$ sets the p-n junction radius, $r_\mathrm{pn}$. (c) Expected `spirograph' motion of carriers in the vicinity of the potential well (not to scale). A Hall field ($E_\mathrm{H}$) points downwards, and carriers drift with velocity $v_\mathrm{d}$ from left to right via cyclotron motion along equipotential (solid black) lines. Incident carriers divert around potential well at a radius $r_d > r_\mathrm{pn}$ (solid grey line) determined by $U(r)$. (d)-(e) Example STP measurements of $\mu_\mathrm{ec}$ around an electrostatic barrier at two magnetic fields ($B = 0$ and 0.4 T) and fixed current ($I_\mathrm{sd} = 200$ $\mu$A and $120$ $\mu$A, respectively). The scale bars are 250 nm.  (d) and (e) are representative images from different experimental runs.}\label{Scheme}
\end{figure}

A schematic of our STP measurement is shown in Fig. \ref{Scheme}a. A current bias ($I_\mathrm{sd}$) is applied across two ends of a graphene/hBN device, and the resulting electrochemical potential ($\mu_\mathrm{ec}= \phi - \mu / e$) variation across the surface is measured in millivolts with a sharp STM tip, where $\phi$ and $\mu$ are the local electrostatic and chemical potentials, respectively, and $e$ is the electron charge. The value of $\mu_{ec}$ at each point is defined by the sample bias ($V_\mathrm{s}$) required to zero the tunneling current ($I_\mathrm{t}$), which we determine by recording an $I_\mathrm{t}$-$V_\mathrm{s}$ curve and performing a linear fit. To characterize the transport quality and cleanliness of our device, we analyzed STP data taken in a large (1.5$\times$1.5 $\mu$m), pristine area of the graphene sample with and without an applied magnetic field. At $B = 0$ T in Fig. \ref{Scheme}a, $\mu_{ec}$ varies linearly between the source and drain electrodes. Assuming a constant current density throughout the device, an application of the Drude conductivity formula yields $l_\mathrm{mr} > 2$ $\mu$m \cite{horng2011}. In a finite magnetic field, the profile of $\mu_{ec}$ resembles a plane the orientation and slope determined by the local Hall angle and Hall field strength, respectively, as shown in Fig. \ref{Scheme}a. We find that the measured Hall angle is nearly 90 degrees at the relatively low field of 0.6 T, which further evidences the long mean free paths in our device. We fit the Hall angle as a function of magnetic field using Ohmic transport theory and obtain a separate estimate of $l_\mathrm{mr}$, also approximately 2 $\mu$m (see Supplemental Information). 

In order to probe magnetotransport around a strong potential perturbation, we introduced a circular electrostatic barrier to the scan area of Fig. \ref{Scheme}a by using the STM tip to ionize defects in the  hBN substrate \cite{lee2016}, creating a screened potential in the overlying graphene. This potential well forms a circular p-n junction in the graphene sheet with a gate-tunable radius ($r_\mathrm{pn}$) ranging from 250 nm ($V_\mathrm{g} = -40$ V) to 400 nm ($V_\mathrm{g} = -2$ V). Previous STM experiments have studied the electronic properties of these p-n junctions, finding a rich interplay between quantum confinement of Dirac fermions and magnetic field-induced interaction effects \cite{lee2016, gutierrez2018}. In Fig. \ref{Scheme}b, we present $dI_\mathrm{t}/dV_\mathrm{s}$ spectra at $B = 1.4$ T and $V_\mathrm{g} = -30$ V as a function of radial distance from the center of the potential well ($r = 0$). At large radii, the tunneling spectra reveal peaks that match the massless, hole-like Landau level (LL) energies $E_N = -\hbar \omega_c \sqrt{N}$, where $\hbar \omega_c = \sqrt{2e\hbar v_F^2 B}$ is the cyclotron energy, $\hbar$ is the reduced Planck constant, $v_\mathrm{F}$ is the Fermi velocity of graphene, and $N$ is a positive integer (see Supplementary Information). For decreasing radii approaching the p-n boundary, the LL peaks shift downwards in energy due to band bending along the potential well profile, $U(r)$. Inside the p-n junction, a different set of peaks match the electron-like LL energies that also shift with the confining potential. Similar band bending of LLs across a p-n junction was first reported by Gutierrez et al. \cite{gutierrez2018}. We obtain an estimate of $U(r)$ -- used later in transport calculations -- by tracking the energy of the first LL as a function of radius (see Fig. \ref{Scheme}b and Supplementary Information).   

Band bending of LLs is also known to occur in response to a Hall field \cite{yang2002, greenaway2021}. For high mobility samples, the Hall potential removes LL degeneracy and all single-particle wavefunctions resemble cyclotron orbits that drift along equipotential lines, as shown in Fig. \ref{Scheme}c. The equipotential lines bend under the p-n junction, just outside a radius $r_\mathrm{d} > r_\mathrm{pn}$ where the potential overcomes the Hall field, $\vert \vec \nabla U(r_d) \vert \approx E_\mathrm{H}$ (grey circle in Fig. \ref{Scheme}c). Neglecting spin and valley degrees of freedom, each of the the color-coded trajectories in Fig. \ref{Scheme}c is uniquely specified by a LL index $N$ and momentum $k$, corresponding to scattering states $\psi_{N,k}(\vec r)$ of the massless Dirac equation in a magnetic field with energies $E_{N,k}$. A non-equilibrium population of these scattering states under transport conditions induces a local charge pile-up, which in turn modifies $\mu_\mathrm{ec}$. In the linear-response regime, assuming the STM tunneling matrix element is constant and ignoring any tip gating and thermovoltage effects, \cite{chu1990, zwerger1991}
\begin{equation}
    \mu_\mathrm{ec}(\vec r) = \frac{1}{e g(\vec r)} \sum_{N, k} \vert \psi_{N, k} (\vec r) \vert^2 \delta f_{N,k} 
\end{equation}
where $g(\vec r)$ is the local thermodynamic density of states at the Fermi level and $\delta f_{N,k}$ is the deviation of the distribution function away from equilibrium, which is defined as follows. Each state $(N, k)$ is injected into the system from a particular contact (the source or the drain) with an electrochemical potential $E_\mathrm{c} = E_s$ or $E_d$, respectively. The change in the electron distribution function is $\delta f_{N, k} = f_0(E_{N, k} - E_F) - f_0(E_{N, k} - E_c)$ where $f_0(E)$ is the Fermi-Dirac distribution. Equation (1) is valid in the ballistic transport regime, when carrier momentum $k$ is a long-lived quantity. Calculations using a slight variation of Eq. (1) in zero magnetic field provide a fully quantum treatment of the Landauer residual-resistivity dipole (LRRD) observed in STP measurements \cite{zwerger1991, briner1996, krebs2023}. To our knowledge, the utility of Eq. (1) for modeling ballistic flow in finite magnetic fields has not yet been tested in an experiment. Alternatively, the Boltzmann equation can be solved in the presence of a magnetic field in certain special cases \cite{holder2019, raichev2020}. Either way, we expect a breakdown of Ohm's law on measurement length scales comparable to the cyclotron radius $R_c$. 

\begin{figure}[H]
\centering
\includegraphics[width=\textwidth,keepaspectratio]{"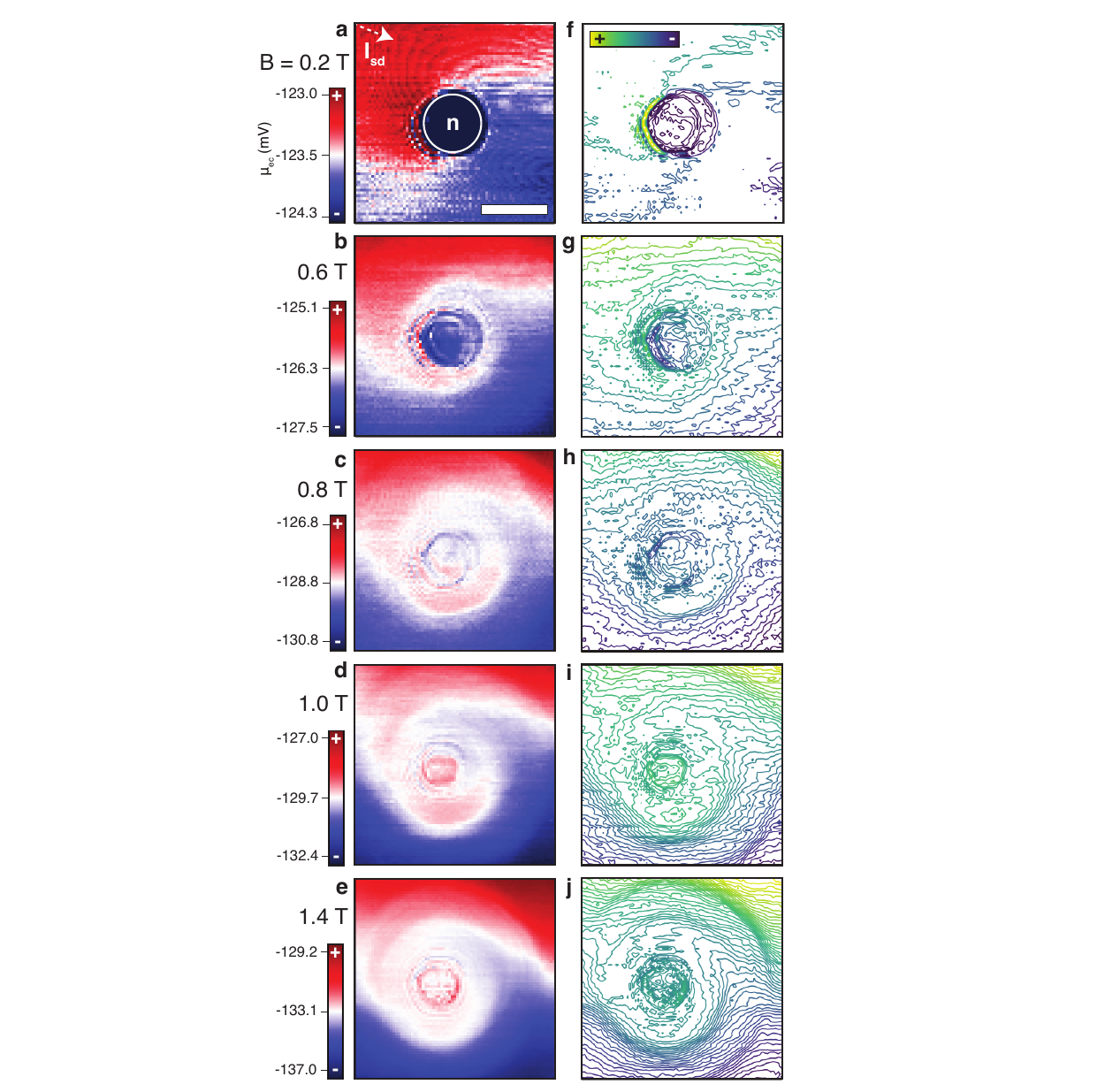"}
\caption{(a)-(f) STP measurements of $\mu_\mathrm{ec}$ around an electrostatic barrier at fixed hole doping (V$_\mathrm{g}$ = -38 V) and current ($I_\mathrm{sd} = 100$ $\mu$A) at six selected magnetic fields between 0 and 1.4 T. The dashed white arrow indicates the direction of incident current flow (inferred from the LRRD in Fig. \ref{Scheme}), and the circular solid white line marks the p-n boundary. The scale bar is 500 nm. (g)-(l) Equipotential contour plots of each STP map in (a)-(f).}\label{Data}
\end{figure}

Characteristic STP images captured after the formation of a circular p-n junction are shown in Fig. \ref{Scheme}d,e at $B = 0$ T and 0.4 T. In zero magnetic field, $\mu_\mathrm{ec}$ is locally higher and lower on opposite sides of the circular p-n barrier, resembling a dipole potential seen in previous STP measurements of graphene \cite{krebs2023}. We identify this feature as an LRRD formed by the backscattering of carriers at the p-n interface, which leads to the accumulation (depletion) of charge on the upstream (downstream) sides of the n-doped region. The LRRD orientation is aligned with the direction of incident current flow (white dashed arrow) \cite{landauer1957}. At $B = 0.4$ T, $\mu_\mathrm{ec}$  exhibits a spiral contour that wraps around a dark blue, circular feature in the center. Like in Fig. \ref{Scheme}d, the edge of this dark circle coincides with the p-n boundary, where sharp variations in the local density of states (LDOS) generate significant thermovoltages between the tip and sample \cite{stovneng1990, krebs2023}. In all STP measurements we observe a small thermovoltage-induced variation of $\mu_{ec}$ at, and within, the p-n boundary; the focus in this work is on the properties of $\mu_{ec}$ outside this boundary.

In Fig. \ref{Data} we report more detailed STP data in magnetic fields varying from $B = 0.2$ to 1.4 T, and for large hole doping ($V_\mathrm{g}$ = -38 V) when the p-n junction radius is relatively small ($r_\mathrm{pn} \approx 250$ nm). At low magnetic fields, $B \leq 0.6$ T (Fig. \ref{Data}a-c), the evolution of $\mu_\mathrm{ec}$ is similar to that observed in Fig. \ref{Scheme}. As the magnetic field increases the equipotential contours of $\mu_\mathrm{ec}$ begin to spiral counter-clockwise around the boundary of the junction, twisting $\sim 90^{\circ}$ at $B = 0.2$ T and beyond $180^{\circ}$ at $0.6$ T. At higher fields, $B > 0.6$ T, the spiral-like patterns persist in $\mu_\mathrm{ec}$, but new features emerge. First, rather than terminating at the p-n boundary, the spiral arms begin to converge at a slightly larger radius of $\sim 400$ nm. This feature can be identified most clearly by the a bunching of equipotential lines along the disk edge in high fields (Fig. \ref{Data}i-j). Second, another disk-like feature of still larger radius (700 - 800 nm) appears in $\mu_\mathrm{ec}$, which also can be most directly visualized in the equipotential contour plots. Both features become more apparent at larger magnetic fields. The inner disk radius shows little dependence on applied field, while the outer disk radius decreases as the field is increased and gate voltage is decreased (see Supplementary Information for gate dependent data). In the remainder of this report, we discuss how these features in $\mu_\mathrm{ec}$ reveal properties related to diffusive and ballistic transport.
 
\begin{figure}[H]
\centering
\includegraphics[width=\textwidth,keepaspectratio]{"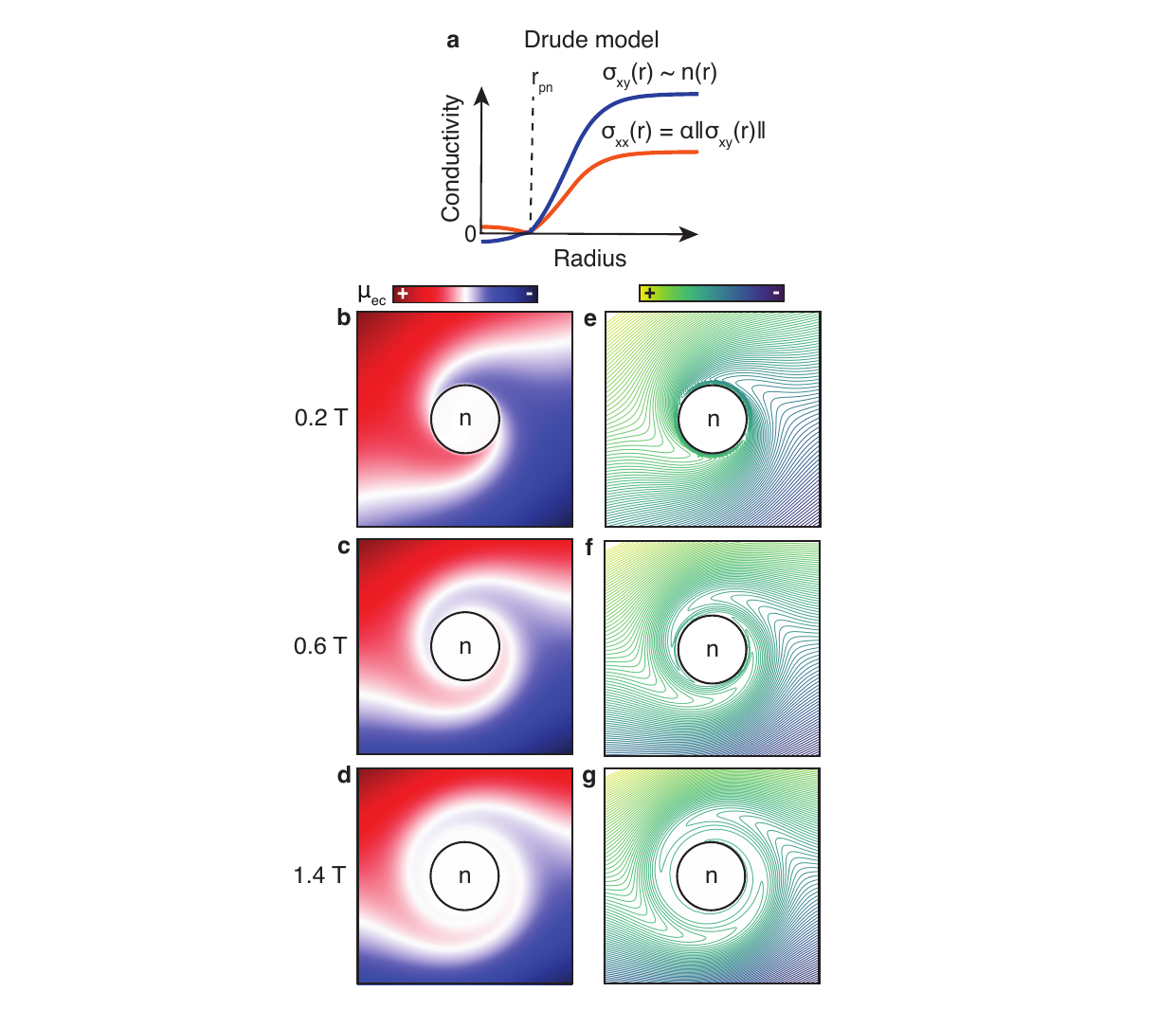"}
\caption{Radial dependence of the longitudnal and Hall Drude conductivities for the measured potential profile U(r). (b)-(d) Current flow-induced electrochemical potential in the Ohmic transport regime, solving Eq. (2) at large hole doping ($E_F = -210$ meV) and applied fields (a) B = 0.2 T ($\alpha = 0.380$), (b) B = 0.6 T ($\alpha = 0.130$), and (c) B = 1.4 T ($\alpha = 0.053$). The black solid line denotes the p-n boundary with a radius of 250 nm. The flow of current in the absence of the barrier runs horizontally from left to right. (d)-(f) Equipotential contours corresponding to the color maps in (a)-(c). }\label{Ohmic}
\end{figure}

At a qualitative level, we can explain the spiral-like patterns observed in low fields by modeling the p-n junction as a hard wall scatterer (vanishing conductivity for $r < r_\mathrm{pn}$), since Dirac fermions cannot easily transmit through p-n junctions in a magnetic field \cite{shytov2007}. In this case, we obtain an exact Ohmic solution -- to Eq. (2) below with constant $\sigma_{xx}, \sigma_{xy}$ -- by adding a pure 2D dipole potential to that of an orthogonal uniform Hall field \cite{landauer1978, zwerger1995}. This solution, however, cannot produce spiral-like contours with $>90^{\circ}$ twist angle, and it also cannot reproduce the disk-like features observed at higher fields. An improved model should take into account the true potential profile $U(r)$, which we measure by tracking the local band bending of LLs (as shown in  Fig. \ref{Scheme}b). In order to numerically calculate $\mu_\mathrm{ec}$, we assume that Ohm's law holds locally, $\vec j (\vec r) = \hat \sigma (\vec r) \vec E(\vec r)$, where $\vec j$ is the current density, $\hat \sigma$ is the conductivity tensor, and $\vec E = -\vec \nabla \mu_\mathrm{ec}$ is the transport-induced electric field. Combining Ohm's law with local charge conservation, $\vec \nabla \cdot \vec j(\vec r) = 0$, we find a second-order partial differential equation for $\mu_\mathrm{ec}$:
\begin{equation}
\sigma_{xx} \nabla^2 \mu_\mathrm{ec} + \vec \nabla \mu_\mathrm{ec} \cdot (\vec \nabla \sigma_{xx} + \hat z \times \vec \nabla \sigma_{xy} ) = 0
\end{equation}
According to the Drude theory, the longitudinal ($\sigma_{xx}$) and Hall ($\sigma_{xy}$) conductivities in a strong magnetic field are $\sigma_\mathrm{xy}(r) \propto n(r) / B$ and $\sigma_{xx} = \alpha \sigma_{xy}$, where the local charge density $n(r) \propto U^2(r)$ varies according the electrostatic barrier profile $U(r)$, and $\alpha = 1/(\omega_\mathrm{c} \tau_\mathrm{mr}) = R_c / l_\mathrm{mr} \ll 1$. Here, $\hbar \omega_c = e B v_F / (k_F c)$ is the cyclotron energy gap at the Fermi level and $\tau_\mathrm{mr} = l_\mathrm{mr} / v_F$ is the momentum-relaxing scattering time. We experimentally determine the former by measuring the LL peak spacing in the tunneling $dI/dV$ data, and the latter from previous STP maps ($\tau_\mathrm{mr} = 2$ $\mu$m$/ v_\mathrm{F} \approx 2$ ps). Figure \ref{Ohmic}a depicts the spatial dependence of the Drude conductivities used in our calculations. Solving Eq. (2) numerically with the corresponding values of $\alpha$, we obtain the solutions shown in Fig. \ref{Ohmic}. For modestly small values of $\alpha$, corresponding to intermediate magnetic fields (Fig. \ref{Ohmic}b,c), the potential takes the form of a spiral that winds tightly around the p-n junction in agreement with the data in Fig. \ref{Data}a,b. For very small values of $\alpha$, corresponding to strong magnetic fields (Fig. \ref{Ohmic}d), the boundary of the spiral arms are shifted away from the p-n junction edge by a small distance (around 70 nm at $B = 1.4$ T). This outward displacement is an example of a general principle for inhomogeneous 2D conductors satisfying $\sigma_{xy} \gg \sigma_{xx}$ \cite{ruzin1993}: current flow tends to concentrate along lines of constant charge density (Hall conductivity), avoiding any depletion regions. In the presence of a circular depletion region with a finite decay length into the bulk, such as the electrostatic barriers in this experiment, current diverts around the depletion region at progressively larger radii as $\alpha$ decreases. Thus, Eq. (2) predicts a radial `turning point' for current flow greater than $r_\mathrm{pn}$ when $\alpha$ is small, lending a possible explanation for the inner disk features in Fig. 2. We note, however, that the predicted radii of these turning points are smaller than the  experimentally observed inner disk radii in the data. Moreover, the Ohmic simulations fail to reproduce the larger outer disk feature observed in our data at high fields. These differences can be seeing most clearly by comparing the contour maps in Fig. \ref{Data}j and \ref{Ohmic}g. In a more accurate model of transport, the semiclassical motion of particles must be considered, along with local changes in the conductivity owing to LL quantization. In particular, partial energy gaps open between adjacent LLs which -- for our doping profile -- would form weakly conducting circular bands that block current flow at a radius greater than $r_\mathrm{pn}$. This motivates further modeling in a way that includes ballistic, quantum, and nonlocal effects.

\begin{figure}[H]
\centering
\includegraphics[width=12cm,keepaspectratio]{"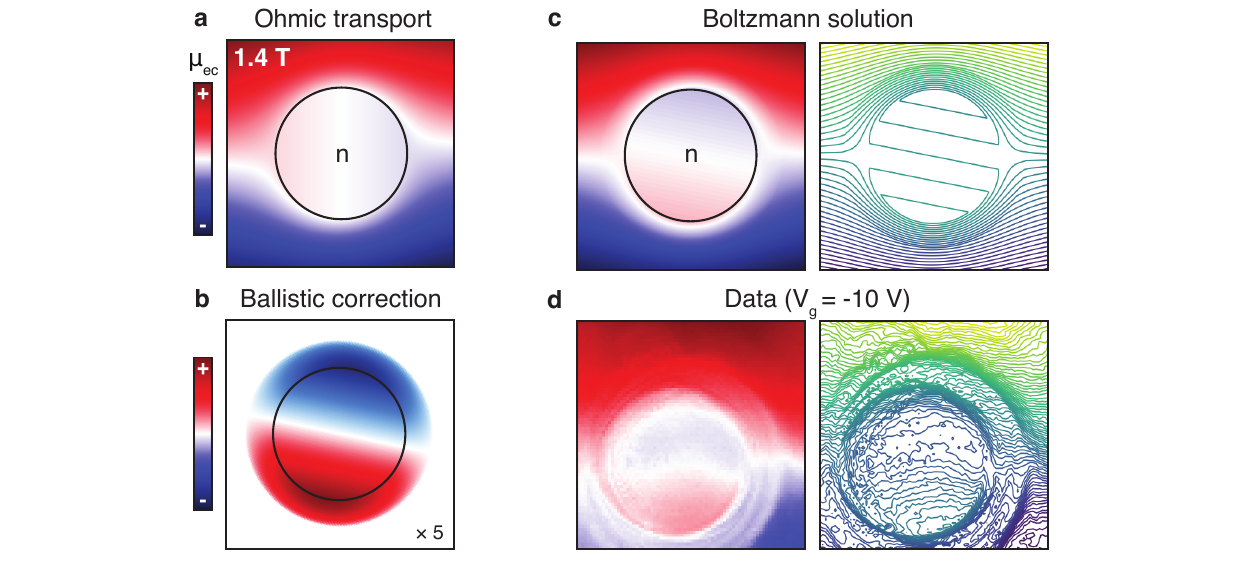"}
\caption{ (a) Electrochemical potential for Ohmic transport around a hard wall. (b) Ballistic correction to purely Ohmic transport obtained from a simplified version of the Boltzmann equation, Eq. (3). (c) The full Boltzmann solution, obtained by adding Fig. \ref{Boltzmann}a,b together. (d) STP data recorded at $B = 1.4$ T and $V_\mathrm{g} = -10$ V.}\label{Boltzmann}
\end{figure}

We first look for corrections by solving the Boltzmann equation 
\begin{equation}
\vec v \cdot \vec \nabla_{\vec r} f_{\vec p}(\vec r) + \big ( e\vec E(\vec r) + e [ \vec v \times \vec B ] \big ) \cdot \vec \nabla_{\vec p} f_{\vec p} (\vec r) = I_{\vec p}(\vec r)
\label{eq:boltzmann}
\end{equation}
where $f_{\vec p} (\vec r)$ is the distribution function of carriers depending on both particle position $\vec r$ and momentum $\vec p$, $\vec E$ is the self-consistent electric field acting on the charge carriers, $\vec v = v_F \hat p$ is velocity of graphene's massless Dirac fermions, and $I_{\vec p} (\vec r)$ is the collision integral describing all relevant scattering processes. A rigorous solution can only be obtained for the hard wall model -- that is, if we approximate $U(r)$ by an infinite step-like discontinuity, and assume a vanishing conductivity for $r < r_\mathrm{pn}$. In this case, the tools in Refs. \cite{holder2019, raichev2020} that apply to the hard wall edges of a conducting channel can be mapped onto the problem of a circular barrier (see Supplementary Information). In the limit of $d_\mathrm{c} \ll r_\mathrm{pn}$, the full Boltzmann solution decomposes as a sum of the solution of Eq. 2 and a ballistic correction, as shown in Fig. \ref{Boltzmann}a,b. This correction is non-negligible only within the distance $2R_c$ from the hard wall, at which point it shows a characteristic kink. The overall magnitude of the correction is proportional to the Hall field outside the barrier. As a result, it is approximately dipolar in shape. The full Boltzmann solution in Fig. \ref{Boltzmann}c therefore contains local peaks in the Hall field one cyclotron diameter away from the p-n boundary. Physically, we can interpret this Hall response as a finite size effect resulting from so-called `skipping orbits' of electrons localized near the p-n interface \cite{holder2019, raichev2020}. We do not explicitly solve for the potential inside the barrier and instead construct it by the linear interpolation of the potential just outside the barrier.
In our experiment, the hard wall model can be most closely realized at low carrier densities when $r_\mathrm{pn}$ is relatively large. In Fig. \ref{Boltzmann} we show representative STP data at a low carrier density when $r_\mathrm{pn} \sim 400$ nm, and find excellent agreement with the Boltzmann solution. We also note that Fig. \ref{Boltzmann}c,d bears a strong resemblance to the high field data in Fig. \ref{Data}e,j -- where the influence of $U(r)$ is not negligible far away from the interface -- if we identify the inner and outer disk features with $r_\mathrm{pn}$ and $r_\mathrm{pn} + 2R_c$, respectively. While the experimentally observed inner disk radius is always larger than $r_\mathrm{pn}$, we suspect that the hard wall model used in Fig. \ref{Boltzmann} underestimates the turning point radius caused by full spatial extent of $U(r)$, which has a decaying tail that could deflect incident carriers at a greater distance.

To explore possible quantum effects, we also utilized tight-binding model simulations \cite{groth2014} to calculate the scattering wavefunctions depicted in Fig. \ref{Scheme}c and obtain an estimate of $\mu_\mathrm{ec}$. In these simulations, we include $U(r)$ as a spatially-varying onsite energy, and a uniform magnetic field through a Peierls substitution. To model the effects of carrier scattering with phonons and disorder, we add a random, onsite potential with maximum amplitude $U_d$. Lastly, we add a linear Hall potential across the sample width. Simulations were carried out using the Kwant software package \cite{groth2014} on a 1000 by 800 unit cell honeycomb lattice. The carbon-carbon bond length was scaled up by a factor of 14 and the hopping energy reduced by a factor of 14 in order to keep the computation time reasonably low while preserving the LL spectrum \cite{lu2011}. Ohmic contacts, which are necessary to equilibrate left- and right-moving carriers, are included by increasing the disorder amplitude $U_d$ near the edges (see Supplementary Information).

\begin{figure}[H]
\centering
\includegraphics[width=\textwidth,keepaspectratio]{"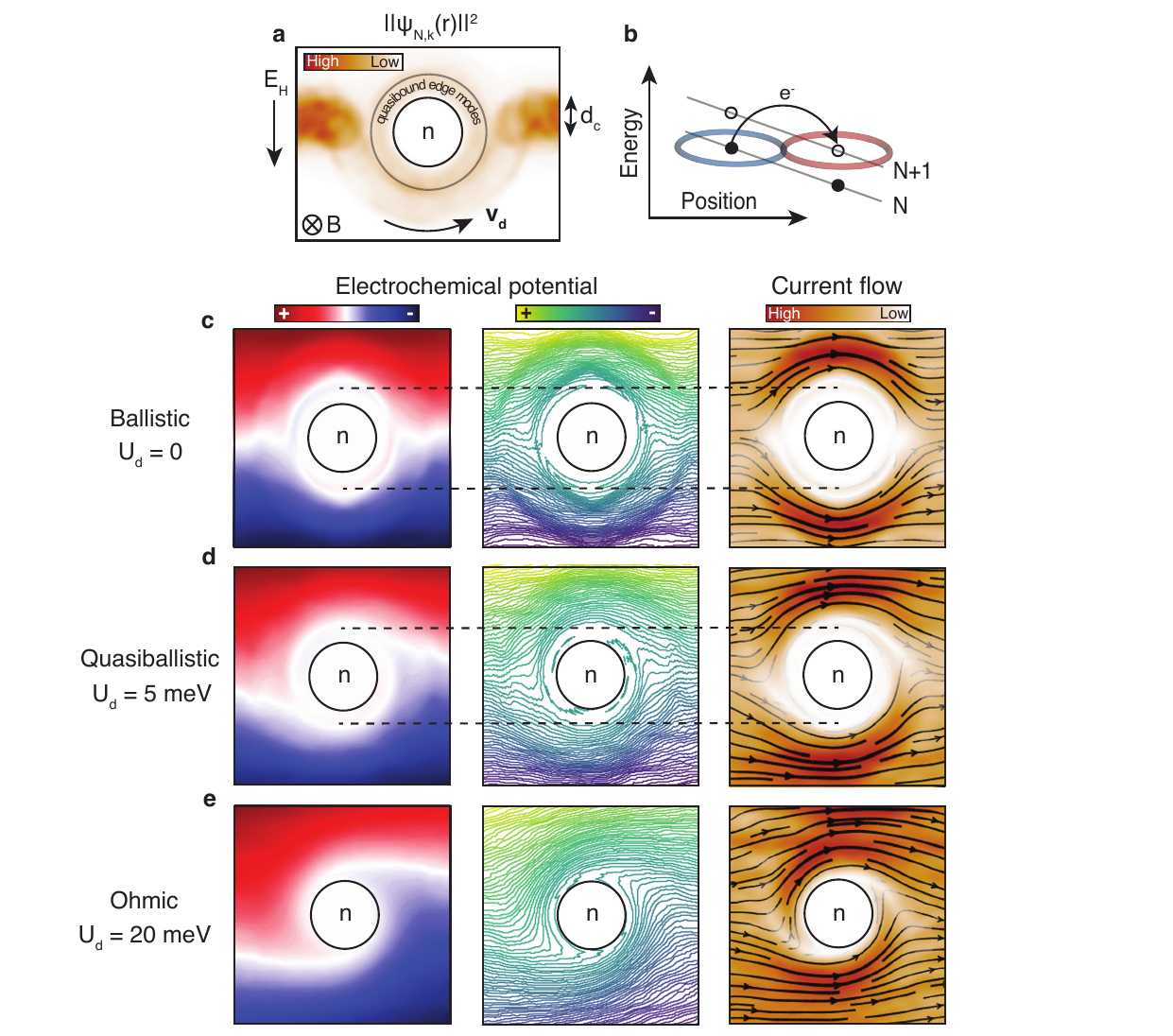"}
\caption{(a) A single tight-binding wavefunction $\vert \psi_{N,k}(\vec r) \vert^2$ with $E_{N,k} = -190$ meV. (b) Schematic of inter-LL scattering processes induced by wavefunction overlap in sufficiently strong Hall fields. An electron in an occupied LL (filled black circle, blue halo) elastically scatters into an adjacent unoccupied LL (hollow circle, red halo).  (c)-(d) Electrochemical potential and current flow of carriers near the Fermi level in the (a) ballistic, (b) quasiballistic, and (c) Ohmic transport regimes.}\label{Kwant}
\end{figure}

An example scattering wavefunction $\vert \psi(E, \vec r) \vert^2 = \vert \psi_{N,k}(\vec r) \vert^2$ for fixed energy $E_{N,k} = -190$ meV (below the Dirac point) at $B = 1$ T is shown in Fig. \ref{Kwant}a. The form of this wavefunction is consistent with the semiclassical picture of 2D ballistic transport in magnetic fields, where the guiding center motion of the carrier traverses the sample along open equipotential contours, as illustrated in Fig. \ref{Scheme}c. The simulations also capture elastic scattering into the aformentioned skipping orbits that propagate along closed equipotential contours just outside the barrier \cite{gutierrez2018}. So far, our model does not include any dissipation. In reality, an electron in the state $\psi_{N,k}$ can interact with phonons or impurities and scatter into a nearby trajectory $\psi_{N,k+q}$ with a guiding center shifted laterally by a distance less than $2R_c$. We approximate such processes by using a large enough Hall field such that adjacent LL wavefunctions overlap, generating inter-LL transitions with a rate modulated by the disorder amplitude $U_d$ (see Fig. \ref{Kwant}b). These inter-LL transitions result in a nonuniform occupation $\delta f_{N,k}$ of LLs, since carriers near the bottom edge of sample are more frequently backscattered into left-moving edge modes. The scattering wavefunction at fixed energy becomes a weighted sum  $\vert \psi(E,\vec r) \vert^2 = \sum \vert \psi_{N,k}(\vec r) \vert^2 \delta f_{N,k}(\vec r)$ over indices $(N,k)$ satisfying $E_{N,k} = -190$ meV. The electrochemical potential can now be obtained by performing a sum $\mu_\mathrm{ec}(\vec r) = \sum_{E\approx E_F} \vert \psi(E,\vec r) \vert^2$ over many such scattering wavefunctions in a narrow energy window around the Fermi level, which coincides with Eq. (1). We note that this methodology is not fully self-consistent, as the scattering wavefunctions used to calculate $\mu_\mathrm{ec}$ are solved in the presence of a linear Hall potential drop instead of the true potential $\mu_\mathrm{ec}$. We expect that any self-consistency corrections will be small, since the potential $U(r)$ is large compared to the variation of $\mu_\mathrm{ec}$ (up to $10$ meV) and should largely dictate the carrier dynamics in its vicinity. Despite the above approximations, these simulations should provide a rough, qualitative prediction of $\mu_\mathrm{ec}$ within the framework of Eq. (1). In Figure \ref{Kwant}c-e we plot the resulting profiles of $\mu_\mathrm{ec}$ for transport conditions that consider different scattering rates, which we tune via $U_d$. In the ballistic transport regime ($U_d = 0$, Fig. \ref{Kwant}c), when there is no carrier scattering away from the Ohmic contacts, $\mu_\mathrm{ec}$ contains two disk features defined by a clear bunching of the equipotential lines. The first, inner disk is larger than the p-n boundary, and corresponds to the radius $r_\mathrm{d}$ at which guiding center trajectories divert around the depletion region in the manner of Fig. \ref{Scheme}c and \ref{Kwant}a. The outer disk corresponds to the inner disk radius plus a cyclotron diameter, $r_d + 2R_c$, in agreement with the Boltzmann theory for a hard wall scatterer if we set $r_\mathrm{pn} \approx r_d$. In the quasiballistic regime ($U_d = 5$ meV, Fig. \ref{Kwant}d), when the carrier mean free path is finite but greater than the cyclotron diameter, $l_\mathrm{mr} > 2R_c$, the inner and outer disks are still present while the equipotentials slightly twist around the inner disk. In the Ohmic regime ($U_d = 20$ meV, Fig. \ref{Kwant}e), the disk features are completely destroyed and replaced by a spiral similar to those in Fig. \ref{Ohmic}. Beyond $\mu_\mathrm{ec}$, we also generate the current flow profiles for carriers near the Fermi level in each transport regime by applying the local current operator to the tight-binding wavefunctions. 

\begin{figure}[H]
\centering
\includegraphics[width=\textwidth,keepaspectratio]{"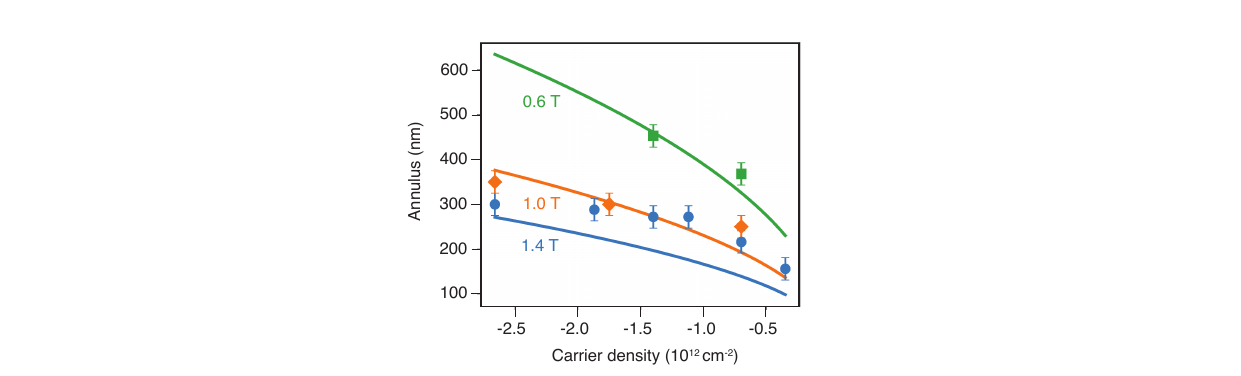"}
\caption{ Measured cyclotron diameter $2R_c$ (outer disk radius minus inner disk radius, or annulus) as a function of carrier density and magnetic field. }\label{Annulus}
\end{figure}

These calculations provide intuition for understanding the field-dependent changes that we observe experimentally (Fig. \ref{Data}). At low magnetic fields, in the regime where $l_\mathrm{mr} < 2R_c$, LL quantization effects are negligible and finite size effects related to the cyclotron radius are obscured by frequent scattering. In this case, the Ohmic transport theory of Eq. (2) is a valid description. At higher fields, when carriers can complete multiple cyclotron orbits before scattering, $l_\mathrm{mr} \gt 2R_c$, changes to the local conductivity due to LL quantization must be taken into account -- even for a partial opening of LL energy gaps -- and nonlocal enhancements of the Hall field begin to develop. We understand the emergence of the inner disk with increasing magnetic field as a consequence of LL quantization, whereby the cyclotron energy gap just below the Fermi level grows larger than the disorder energy scale set by $U_d$, so that states within the disk $r < r_d$ become highly localized and carrier scattering across the boundary at $r_d$ is suppressed. The presence of this effective `hard wall' at $r_d \geq r_\mathrm{pn}$ causes charge carriers to pile-up as they divert around the barrier, resulting in a nonlocal enhancement of the Hall field at $r_d + 2R_c$. 
Measuring the distance between the inner and outer disk features in the data (defining an annulus), we find reasonable agreement with the expected value $2R_c = 2\hbar \sqrt{\pi n} / e B$ across all selected carrier densities and magnetic fields in Fig. \ref{Annulus}. 

In summary, we have used STP to probe magnetotransport in graphene at $T = 4.5$ K with the help of electrostatic barriers `drawn' with the STM tip. As the magnetic field is increased up to 1.4 T, the current flow evolves in the following way: at weak fields, and on measurement length scales less than $l_\mathrm{mr}$, transport is ballistic as evidenced by the LRRD in Fig. \ref{Scheme}d. In intermediate fields such that $l_\mathrm{mr} < 2R_c$, the electron flow is a combination of the lateral diffusion with a slow guiding center drift of the cyclotron orbits. This advection-diffusion transport creates spiral-like current flow around the obstacle. At higher magnetic fields, when $2R_c < l_\mathrm{mr}$, the flow crosses over into a `quasiballistic' regime marked by a local enhancement of the Hall field one cyclotron diameter away from the barrier. Direct measurements of $R_c$ in quantum materials are rare, and we view our technique as complimentary to established magnetic focusing methods \cite{taychatanapat2013, bhandari2016, lee2016}. Additionally, LL quantization begins to suppress the transmission of carriers across an inner disk radius $r_d > r_\mathrm{pn}$ determined by the form of the electrostatic potential $U(r)$. We believe our measurements provide definitive insights into the nature of compressible electron fluid flow in a magnetic field around strong disorder potentials. Single-particle physics is largely sufficient to describe the data in this experiment, although future STP measurements could investigate the role of electron-electron interactions, which are expected to be pronounced at even higher fields in the integer or fractional quantum Hall regimes. 
 
\section*{Acknowledgments}
Work by Z. J. K. and K. J. S. was supported by the U.S. Department of Energy (DOE), Office of Science, Basic Energy Sciences (BES) Program for Materials and Chemistry Research in Quantum Information Science under Award No. DE-SC0020313. W. A. B. and M. A. F. were supported by Office of Naval Research award N00014-20-1-2356. V. W. B. was supported by NSF CAREER award \#239478. The authors gratefully acknowledge the use of facilities and instrumentation supported by NSF through the University of Wisconsin Materials Research Science and Engineering Center (No. DMR1720415). K. W. and T. T. acknowledge support from the JSPS KAKENHI (Grant Numbers 21H05233 and 23H02052) and World Premier International Research Center Initiative (WPI), MEXT, Japan. M.A.F. was also provided through a Google PhD Fellowship (Quantum Computing).
 
\bibliography{sn-bibliography}
\end{document}